\documentclass[review]{elsarticle}

\usepackage{hyperref}
\usepackage{graphicx}
\usepackage{natbib}

\begin{document}

\begin{frontmatter}

\title{Low-friction, wear-resistant, and electrically homogeneous multilayer graphene grown by chemical vapor deposition on molybdenum}

\author[IPB]{Borislav Vasi\'c\corref{mycorrespondingauthor}}
\cortext[mycorrespondingauthor]{Corresponding author}
\ead{bvasic@ipb.ac.rs}

\author[IPB]{Uro\v s Ralevi\'c}

\author[IHTM]{Katarina Cvetanovi\'c Zobenica}

\author[IHTM]{Mil\v ce M. Smiljani\'c}

\author[IPB]{Rado\v s Gaji\'c}

\author[IHTM]{Marko Spasenovi\'c}

\author[DELFT]{Sten Vollebregt}

\address[IPB]{Graphene Laboratory of Center for Solid State Physics and New Materials, Institute of Physics Belgrade, University of Belgrade, Pregrevica 118, 11080 Belgrade, Serbia}
\address[IHTM]{Center  of  Microelectronic  Technologies,  Institute  of  Chemistry,  Technology  and  Metallurgy, University of Belgrade, Njego\v seva 12, 11000 Belgrade, Serbia}
\address[DELFT]{Department of Microelectronics, Delft University of Technology, Feldmannweg 17, 2628CT Delft, The Netherlands}

\begin{abstract}
Chemical vapour deposition (CVD) is a promising method for producing large-scale graphene (Gr). Nevertheless, microscopic inhomogeneity of Gr grown on traditional metal substrates such as copper or nickel results in a spatial variation of Gr properties due to long wrinkles formed when the metal substrate shrinks during the cooling part of the production cycle. Recently, molybdenum (Mo) has emerged as an alternative substrate for CVD growth of Gr, mainly due to a better matching of the thermal expansion coefficient of the substrate and Gr. We investigate the quality of multilayer Gr grown on Mo and the relation between Gr morphology and nanoscale mechanical and electrical properties, and spatial homogeneity of these parameters. With atomic force microscopy (AFM) based scratching, Kelvin probe force microscopy, and conductive AFM, we measure friction and wear, surface potential, and local conductivity, respectively. We find that Gr grown on Mo is free of large wrinkles that are common with growth on other metals, although it contains a dense network of small wrinkles. We demonstrate that as a result of this unique and favorable morphology, the Gr studied here has low friction, high wear resistance, and excellent homogeneity of electrical surface potential and conductivity.
\footnote{\textcopyright 2019. This manuscript version is made available under the CC-BY-NC-ND 4.0 license http://creativecommons.org/licenses/by-nc-nd/4.0/}
\end{abstract}

\begin{keyword}
graphene, chemical vapour deposition, atomic force microscopy, friction, wear, electrical properties
\end{keyword}

\end{frontmatter}


\section{Introduction}

Chemical vapour deposition (CVD) is the most dominant method for fabrication of large-area single- and few-layer graphene (Gr) films on various metallic substrates (catalysts) \cite{Li_Science, Obraztsov, zhou_nat_comm, Kim_Nature}. After the growth, Gr films are transferred onto desired substrates that allow practical use \cite{Li_NL, transfer_nanoscale}. The main advantages of CVD over other fabrication techniques are its relative simplicity, low cost, and industrial applicability \cite{review_jmchem, review_chemres, review_carbon}. Still, CVD growth and Gr transfer yield films with defects such as grain boundaries \cite{RuizVargas_NL, Lee_Science, Rasool_NatCom, Koepke_NL, Clark_NL, NemesIncze_carbon, Roche_AdvMat}, wrinkles \cite{Xu_NL, Liu_NanoResearch, Ahmad_Nanotech, Avouris_NL, Ladak, wrinkles_kpfm, vasic_cvd_gr_carbon} and cracks. Formation of wrinkles, for example, occurs due to a large difference in thermal expansion coefficients of Gr and the catalytic substrate, which results in different shrinking rates during cooling at the end of the CVD growth process. Wrinkles in Gr have been shown to be highly detrimental to the mechanical robustness and electrical homogeneity of graphene \cite{vasic_cvd_gr_carbon}. CVD graphene is prone to formation of defects both on most commonly used catalytic metal substrates such as copper \cite{Li_Science, Obraztsov, zhou_nat_comm} and nickel \cite{Kim_Nature}, as well as on less traditional substrates such as ruthenium, iridium, and platinum \cite{review_carbon}. 

Recently, Gr grown by CVD on thin molybdenum (Mo) films sputtered on silicon wafers \cite{synthesis_carbon, synthesis_sten, synthesis_NL_paper} has emerged as an alternative to Gr grown on traditional metal substrates. Growth on Mo offers several advantages. Namely, the thermal expansion coefficient of Mo is well matched to that of Gr, supporting “wrinkle-free” growth \cite{synthesis_sten}. Also, Mo has a high melting point, resulting in less restructuring of the Mo substrate compared to copper during the CVD process. Finally, low solubility of carbon in bulk Mo facilitates easy growth of Gr layers \cite{synthesis_carbon}, making the process attractive for high-volume applications. In addition to the advantages for growth, graphene on a thin sputtered layer of Mo has advantages for subsequent processing. The Mo layer can be patterned prior to growth, enabling patterning of CVD graphene without post-growth lithography, and the Mo can be easily removed once Gr is grown on it, which allows transfer-free fabrication of Gr devices that is compatible with CMOS processes \cite{gas_sensor2, pressure_sensing}. Graphene grown with this novel process was shown to have applications in anti-corrosion coatings \cite{corrosion1, corrosion2, oxidation}, gas \cite{gas_sensor2, gas_sensor1} and pressure sensors \cite{pressure_sensing}. Although Gr grown on Mo has high potential for practical use, the relation between microscopic morphology, distribution and geometry of wrinkles, and their influence on the mechanical and electrical properties of the material have not yet been studied.

Using atomic force microscopy (AFM) based methods, here we present nanoscale analysis of the morphology, mechanical and electrical properties of few-layer Gr grown by CVD on Mo. We demonstrate that the material contains very few wrinkles and that those wrinkles have dimensions that are much smaller than those typically encountered in Gr grown on copper foils. Gr grown on Mo has low friction and high wear resistance as demonstrated by friction force microscopy and nanoscale wear tests. Using Kelvin probe force microscopy (KPFM) and conductive AFM (c-AFM) we show that this material has high uniformity of the Fermi level (work function) and electrical conductivity, respectively, over large areas. These results allude to strong potential uses of Gr grown on Mo for both mechanical and electrical applications such as ultrathin solid lubricants, electrodes and membranes for nano and microelectromechanical systems.

\section{Experimental}

\subsection{Graphene fabrication and transfer}

We studied both Gr on Mo as grown, as well as Gr transferred from Mo to $\mathrm{Si/SiO_2}$. The $50 \ \mathrm{nm}$ thick Mo catalyst was deposited using magnetron sputter coating on top of a Si/$\mathrm{SiO_2}$ wafer (p-type, 10~cm, (100) orientation). $\mathrm{SiO_2}$ was approximately 600~nm thick and it was grown using wet thermal oxidation. The Mo target purity was $99.95\%$. Gr was deposited using an AIXTRON Blackmagic Pro system at 915$^\circ$C using 960/40/25 sccm of Ar/H$_\mathrm{2}$/CH$_\mathrm{4}$ at 25 mbar for 30 minutes and cooled to room temperature under an Ar atmosphere. After this, the wafers were cut into smaller dies for sample preparation. Further details of the CVD growth of Gr on Mo can be found in our previous paper \cite{synthesis_sten}.  

Graphene was transferred by first immersing dice of Gr on wafer in 30\% hydrogen-peroxide for 25 minutes. Hydrogen peroxide etches away the Mo underneath the Gr layer and Gr is released, floating on the surface of the hydrogen peroxide solution. Gr was transfered into a Petri dish, 5 cm in diameter and 17 ml of volume, with $\mathrm{H_2O_2}$. Hydrogen peroxide was exchanged with deionized (DI) water. The DI water was exchanged three times to ensure complete removal of peroxide. Gr was then carefully picked up onto a $\mathrm{Si/SiO_2}$ wafer die. The sample was dried at room temperature for 25 minutes, and was put under a glass bell for the next 24 hours to dry completely.

\subsection{AFM and Raman characterization}

All AFM measurements were performed with an NTEGRA Spectra system at ambient conditions. Morphology was measured in tapping AFM mode with NSG01 probes. The surface roughness was measured across ten $50 \times 50 \ \mathrm{\mu m^2}$ areas, calculated as the root-mean square of the height distributions, and then averaged. Phase lag of the AFM probes was measured simultaneously with topography in order to achieve better contrast of small topographic features and to check for possible changes in material contrast on the sample surface.

Wear tests were done in contact AFM mode on 5-10 different $10 \times 10 \ \mathrm{\mu m^2}$ areas with diamond coated probes DCP20. In order to initiate Gr wear, the normal load was kept constant during scanning within $1 \ \mathrm{\mu m}$ wide parallel stripes and increasing in steps of around $1 \ \mathrm{\mu N}$ from stripe to stripe, for a total range between $0.5 \ \mathrm{\mu N}$ and $5 \ \mathrm{\mu N}$ \cite{vasic_cvd_gr_carbon}. During the wear tests, we recorded the lateral forces in both forward and backward directions. These forces correspond to the lateral torsion of the AFM cantilever due to the AFM tip-Gr friction. The friction force was determined as the half-difference between the lateral force in the forward and backward direction. The normal force was calculated according to force-displacement curves, whereas the friction force was calibrated on a standard Si grating \cite{lf_calibraction}.

Kelvin probe force microscopy (KPFM) and Pt coated NSG01/Pt probes were employed in order to measure local electrical surface potential. KPFM is a two-pass AFM-based measurement technique which returns a local contact potential difference (CPD) between a metallic AFM tip and the sample surface. The topographic profile of the sample was measured in the first pass. In the second pass, the AFM probe was lifted by $20 \ \mathrm{nm}$ and scanned along the same topographic line as in the first pass, while a sum of DC and variable AC voltages was applied between the probe and the sample. The role of the AC voltage was to electrically excite probe oscillations, while the DC voltage was controlled by the AFM feedback loop in order to nullify these oscillations. The value of DC voltage which nullifies AFM probe oscillations is equal to the local CPD between the AFM tip and the sample. 

Since CPD is equal to the difference between the work functions of the AFM tip ($\mathrm{WF_{tip}}$) and the sample ($\mathrm{WF_{sample}}$), in order to find the absolute value of $\mathrm{WF_{sample}}$, the $\mathrm{WF_{tip}}$ of the Pt coated AFM probes was calibrated on a HOPG sample with a well known work function of $4.6 \ \mathrm{eV}$ \cite{kpfm_calibration}. Measurements on HOPG yielded $\mathrm{WF_{tip}}=5 \ \mathrm{eV}$. The work functions of the Gr samples were calculated as $\mathrm{WF_{Gr}}=\mathrm{CPD}-\mathrm{WF_{tip}}$. As in the case of the surface roughness and wear tests, the CPD was measured on 5-10 different areas ($50 \times 50 \ \mathrm{\mu m^2}$ in the case of Gr grown on Mo, and $30 \times 30 \ \mathrm{\mu m^2}$ on Gr transferred on $\mathrm{SiO_2}$) and then averaged.

Conductive AFM (C-AFM) with highly doped diamond coated probes DCP20 was used for characterization of local conductivity. In C-AFM, during standard topographic imaging in AFM contact mode, a DC voltage in a range between 1 V and 2 V was applied between the probe and the sample. The resulting DC current through the AFM probe, proportional to the local conductivity of Gr samples, was simultaneously measured with a built-in current amplifier. In order to avoid wear of AFM tips and achieve reliable current measurements, C-AFM was done using diamond coated DCP20 probes which were heavily doped by nitrogen. The diamond coating provides the robustness and wear resistance of AFM tips, while the high doping makes them highly conductive. As a result, these probes enabled reliable current mapping in contact AFM mode. 
 
Raman imaging of CVD Gr transferred on $\mathrm{Si/SiO_2}$ was performed on the same NTEGRA Spectra module equipped with a confocal Raman system (NA 0.7). Raman maps were measured with a step of $0.5 \ \mathrm{\mu m}$. The wavelength of the excitation laser was $532 \ \mathrm{nm}$. 

\section{Results and discussion}

\subsection{Morphology}

\subsubsection{CVD Gr on Mo}

\begin{figure}
	\centerline{\includegraphics[width=8.5cm]{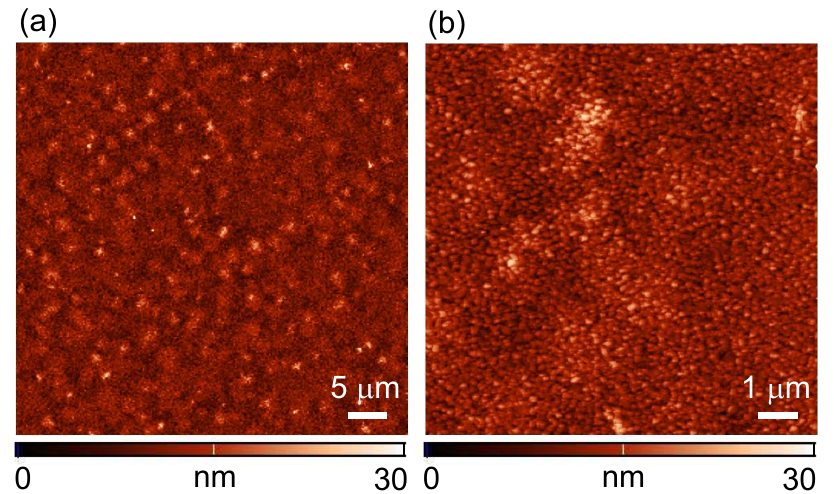}}
	\caption{Morphology of CVD Gr on Mo: (a) $50 \times 50 \ \mathrm{\mu m^2}$ and (b) $10 \times 10 \ \mathrm{\mu m^2}$ area.}
	\label{fig_morphology_on_Mo_1}
\end{figure}

The topography of CVD Gr on Mo is depicted in Fig. \ref{fig_morphology_on_Mo_1} on two different length scales. One of the main motivations and potential benefits of CVD on Mo is the growth of wrinkle-free Gr since the thermal expansion coefficient of Mo is much better matched to that of Gr than the thermal expansion coefficients of copper or nickel. Indeed, in these topographic images there are no long wrinkles typically observed in traditional CVD Gr grown on copper. 

\begin{figure}
	\centerline{\includegraphics[width=8.5cm]{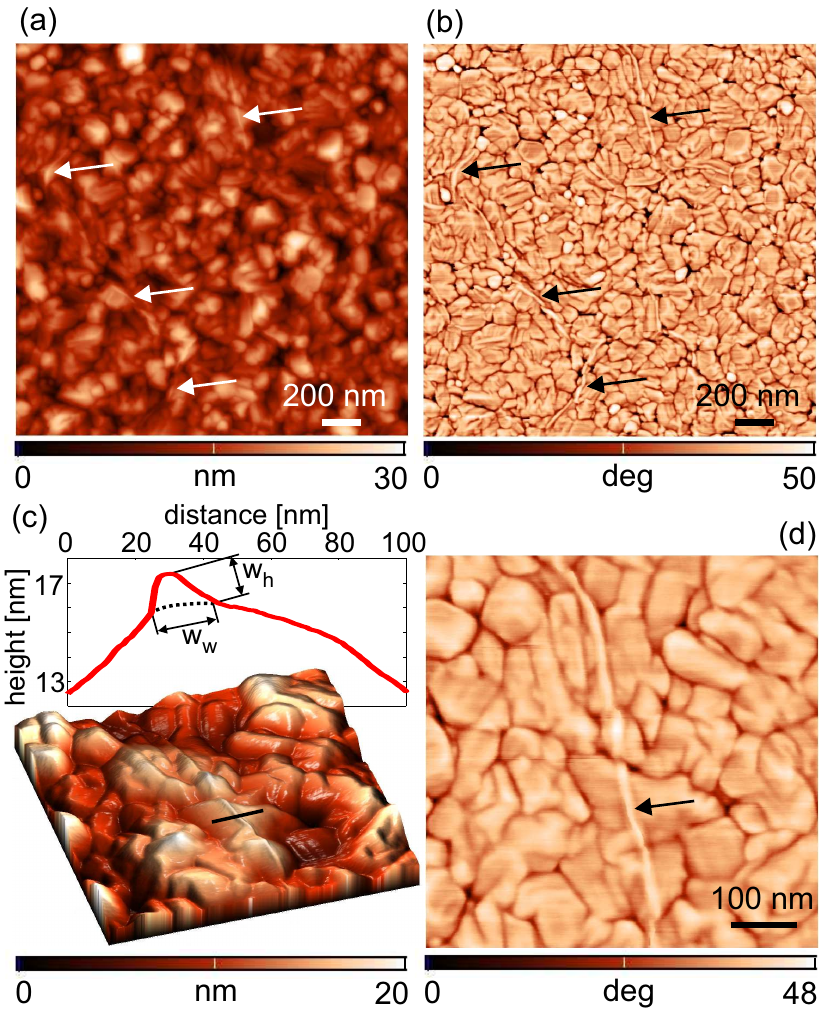}}
	\caption{Wrinkles in CVD Gr on Mo: (a) topography and (b) phase of $2 \times 2 \ \mathrm{\mu m^2}$ area. (c) Three-dimensional topography and (d) phase of $0.6 \times 0.6 \ \mathrm{\mu m^2}$ area. Wrinkles are marked by arrows. The inset in part (c) depicts the topographic cross-section across the wrinkle, along the solid line. The dotted line is a guide to the eye which follows the surface of the grain in order to emphasize the wrinkle geometry.}
	\label{fig_morphology_on_Mo_2}
\end{figure}

Still, very short wrinkles can sometimes be observed on small-scale images. A typical example is presented in the topographic and phase images in Figs. \ref{fig_morphology_on_Mo_2}(a) and \ref{fig_morphology_on_Mo_2}(b), respectively. Usually it is difficult to recognize wrinkles in topographic images since they are very small. On the other hand, wrinkles can be resolved in the phase image as elongated, curved lines, several hundreds of nanometers long (denoted by arrows). By using the position of wrinkles found from the phase image to carefully search the topographic map, one can identify bright and narrow lines indicating that here Gr is locally wrinkled. The local Gr wrinkling is best visualized if we further zoom into an area containing a single wrinkle, as illustrated in Figs. \ref{fig_morphology_on_Mo_2}(c) and \ref{fig_morphology_on_Mo_2}(d), with the three-dimensional topographic and phase image, respectively. The inset in part (c) depicts the height profile across the wrinkle. Its width $\mathrm{w_w}$ and height $\mathrm{h_w}$ are around $20 \ \mathrm{nm}$ and $1.5 \ \mathrm{nm}$, respectively.

Surface roughness of Gr calculated from large-scale images such as from Fig. \ref{fig_morphology_on_Mo_1}(a) was $2.9\pm0.1 \ \mathrm{nm}$ (averaged over ten $50 \times 50 \ \mathrm{\mu m^2}$ areas). Lower roughness could be achieved by using Mo foils as substrates \cite{synthesis_NL_paper}, but foils are not compatible with semiconductor technologies. Therefore, we have considered Gr grown on thin films such as sputtered Mo, which is fully compatible with CMOS processing \cite{gas_sensor2, pressure_sensing}. 

The roughness of Gr originates from patches with slightly increased height and from the grain structure of the underlying substrate. Gr patches in topographic images in Fig. \ref{fig_morphology_on_Mo_1} look like brighter domains, with a lateral size of several microns, and a height of several nanometers. A high-resolution image of an $2 \times 2 \ \mathrm{\mu m^2}$ area, depicted in Fig. \ref{fig_morphology_on_Mo_2}(a), reveals that Gr follows the morphology of the underlying Mo substrate while the grain structure of Mo is imprinted and replicated onto Gr. As a result, the grain structure of Mo dominates the measured topography thus contributing to increased surface roughness. It should be emphasized that as deposited Mo is very flat with sub-nm roughness. However, the Mo turns into $\mathrm{Mo_2C}$ upon exposure to CH$_\mathrm{4}$ before the Gr formation starts \cite{synthesis_NL_paper}, and due to this recrystallization into $\mathrm{Mo_2C}$, the roughness increases. 

Grain structure and especially grain boundaries are even better visualized in the corresponding phase image in Fig. \ref{fig_morphology_on_Mo_2}(b). Since the phase signal is generally very sensitive to sudden and sharp topographic features such as hills (grains) or narrow holes (grain boundaries), it can be used for their visualization with even better resolution than in pure topographic images. As can be seen, the shape of grains is rather irregular, with an average grain diameter of around $100 \ \mathrm{nm}$. The autocorrelation function of the height distribution (not-shown here) is isotropic, indicating that on average, there is no preferential in-plane anisotropy of the grain structure.

\subsubsection{CVD Gr on $\mathrm{SiO_2}$}

\begin{figure*}
	\centerline{\includegraphics[width=8.5cm]{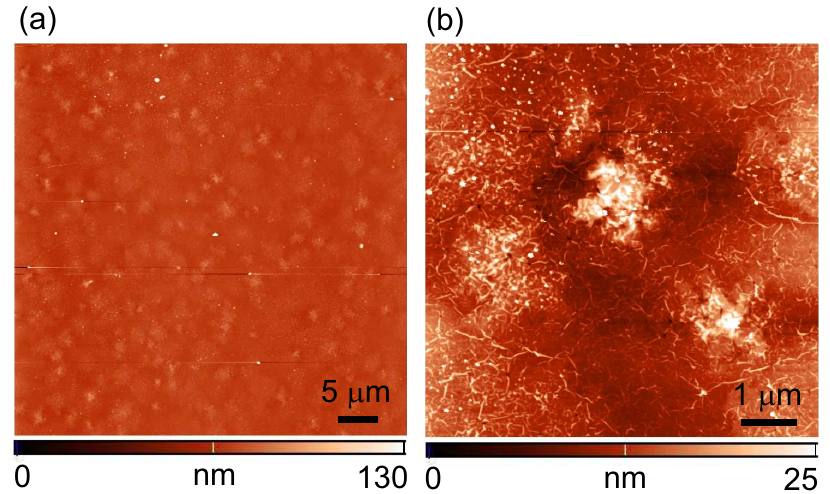}}
	\caption{Morphology of CVD Gr on $\mathrm{SiO_2}$: (a) $50 \times 50 \ \mathrm{\mu m^2}$ and (b) $7 \times 7 \ \mathrm{\mu m^2}$ area.}
	\label{fig_morphology_on_SiO2_1}
\end{figure*}

The morphology of CVD Gr transferred on $\mathrm{SiO_2}$ is presented in Fig. \ref{fig_morphology_on_SiO2_1}. The calculated surface roughness was $3.8\pm1 \ \mathrm{nm}$ (averaged over ten $50 \times 50 \ \mathrm{\mu m^2}$ areas). Therefore, the roughness slightly increased after the transfer mainly because micron-size patches with a slightly increased height were more evident than in the previous case of Gr on Mo. In addition, Gr on $\mathrm{SiO_2}$ has a rather dense network of short wrinkles as depicted in Fig. \ref{fig_morphology_on_SiO2_1}(b) and it contains nano-particles, represented by bright, isolated point-like features in Fig. \ref{fig_morphology_on_SiO2_1}(a), which are most probably residues from the transfer process. 

As mentioned above, Mo thin films as substrates allow transfer-free Gr fabrication, thus making the fabrication simpler and compatible with semiconductor technologies \cite{gas_sensor2, pressure_sensing}. Here we consider Gr transferred on $\mathrm{SiO_2}$ because it facilitates characterization and analysis, but it should be emphasized that the chemical treatment is the same in both cases (i.e. Gr is immersed in hydrogen-peroxide in both cases, the only difference is that in the transfer-free approach, the original growth substrate is reused, whereas during the transfer, another substrate is used for picking Gr up). Therefore, the results presented here hold for transfer-free Gr as well.

\begin{figure*}
\centerline{\includegraphics[width=16cm]{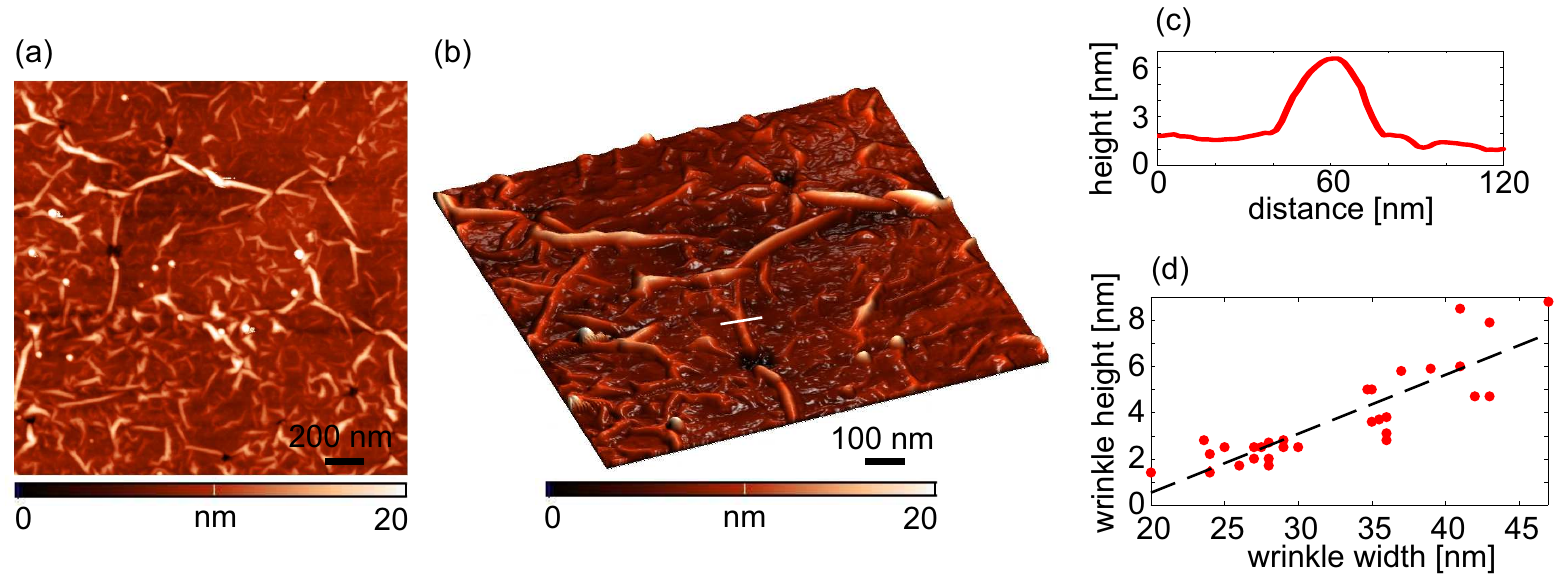}}
\caption{Wrinkles in CVD Gr on $\mathrm{SiO_2}$: (a) two-dimensional topographic image of $2 \times 2 \ \mathrm{\mu m^2}$ area, and (b) three-dimensional $1 \times 1 \ \mathrm{\mu m^2}$ area, (c) the height profile of the wrinkle along the solid line in part (b), and (d) the distribution of wrinkle widths and heights from part (c). The dashed line in the distribution plot is a linear fit.}
\label{fig_morphology_on_SiO2_2}
\end{figure*}

Short Gr wrinkles observed in Fig. \ref{fig_morphology_on_SiO2_1}(b) are better visualized in the small-scale images in Fig. \ref{fig_morphology_on_SiO2_2}(a) and \ref{fig_morphology_on_SiO2_2}(b). The typical height profile of a wrinkle is depicted in Fig. \ref{fig_morphology_on_SiO2_1}(c), whereas the distributions of wrinkle widths and heights are given in Fig. \ref{fig_morphology_on_SiO2_2}(d). The height can be fitted with a linear function of the width. According to Ref. [\citenum{Avouris_NL}], there are three classes of wrinkle geometry: ripples, standing collapsed wrinkles and folded wrinkles. Since the wrinkle width in our case is less than $\sim 50 \ \mathrm{nm}$, they have the geometry of ripples, while higher wrinkles are most probably standing collapsed ones. This is a significant difference compared to CVD Gr grown on copper \cite{vasic_cvd_gr_carbon} where thermally induced wrinkles are much wider, up to several hundreds of nanometers, and belong to the class of folded wrinkles. On the Gr studied here, wrinkles are generally short, most of them with lengths in the range $100-200 \ \mathrm{nm}$. They do not have any preferential direction, while shorter wrinkles are usually curved. These wrinkles appear during transfer onto $\mathrm{SiO_2}$. According to wrinkle lengths, shapes, and their mutual distances, it seems that they correspond to grain boundaries of Gr on Mo - narrow, irregular and curved domains along which Gr on Mo was locally bent and corrugated. The different wrinkle type of Gr grown on Mo compared to that grown on copper is the most probable reason for improved mechanical and electrical properties, as will be discussed in the following sections.

\subsection{Raman analysis}

\begin{figure*}
\centerline{\includegraphics[width=16cm]{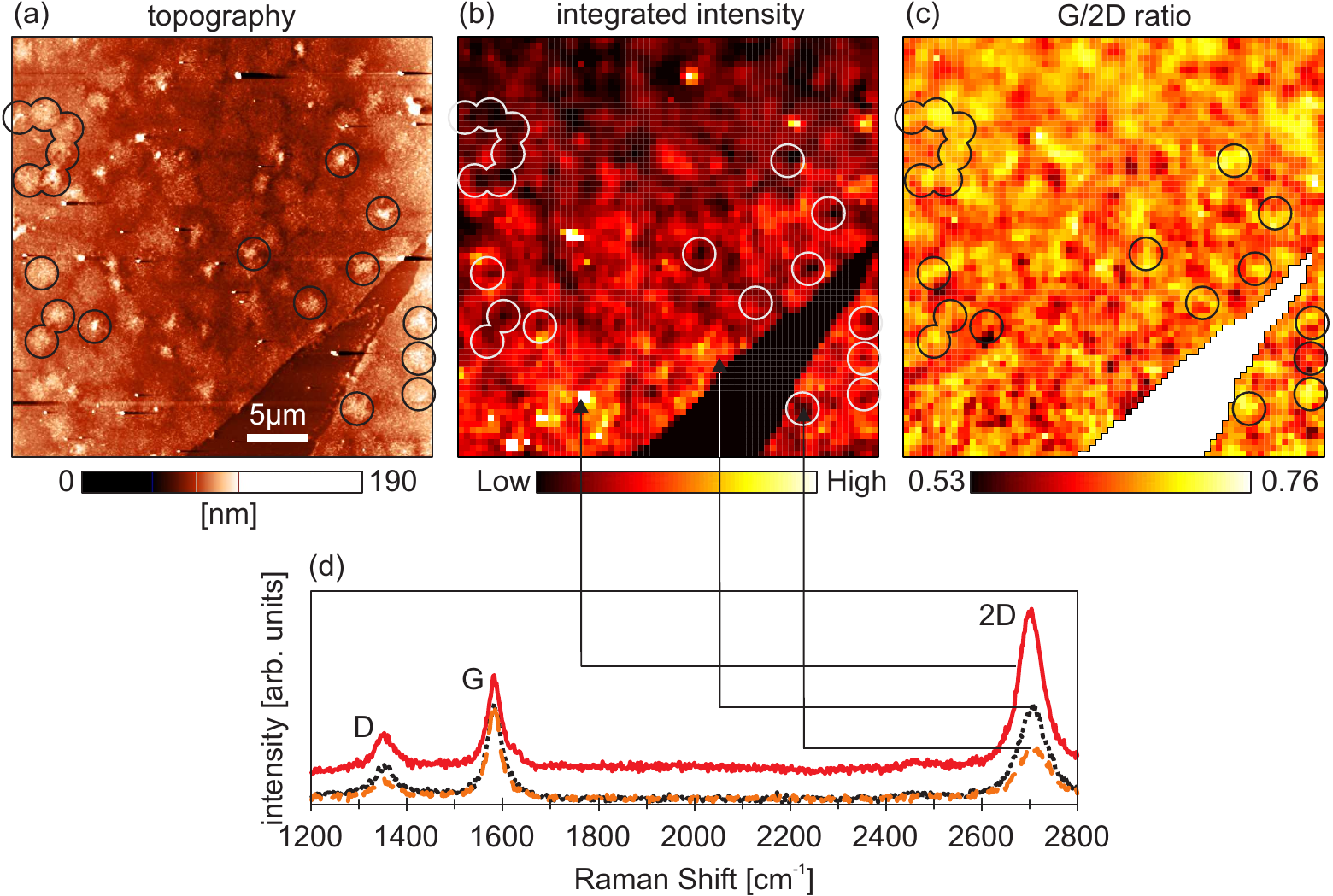}}
\caption{(a) Topography, (b) total Raman intensity integrated between $1200 \ \mathrm{cm^{-1}}$ and $2800 \ \mathrm{cm^{-1}}$, (c) the ratio between the intensity of G and 2D Raman modes, and (d) Raman spectra measured at three points marked in part (b).}
\label{fig_raman}
\end{figure*}

Further characterization of CVD Gr transferred on $\mathrm{SiO_2}$ was done by combined AFM and Raman mapping. The results are presented in Fig. \ref{fig_raman} with the topography (part (a)), integrated Raman intensity (part (b)), the ratio between the intensity of G and 2D Raman modes (part (c)), and Raman spectra taken at three representative points (part (d)). As can be seen, the characteristic Raman modes of Gr, G (around $1586 \ \mathrm{cm^{-1}}$) and 2D (around $2700 \ \mathrm{cm^{-1}}$) modes are clearly resolved. Still, the appearance of the defect mode D (around $1350 \ \mathrm{cm^{-1}}$) indicates non-negligible defects in CVD Gr. The 2D peak is slightly shifted to longer wavenumbers which indicates that the considered CVD Gr is multi-layered. The ratio G/2D is in the range between 0.5-0.7, which corresponds to a thickness of 4-6 layers \cite{graf_nl_20107}. The same thickness is confirmed by AFM measurements as shown in Fig. S1 of Supplementary material. By comparing encircled domains in all maps, most of the patches with increased height correspond to domains with decreased total Raman intensity and increased G/2D ratios. Therefore, the number of Gr layers is locally increased within those patches. Still, some patches where the G/2D ratio is not increased (or has even slightly decreased) likely contain just locally wrinkled and/or folded layers. Raman spectra of samples stored in ambient conditions for over two years reveal no deterioration due to aging. 

\subsection{Friction and wear properties}

\begin{figure}
	\centerline{\includegraphics[width=8.5cm]{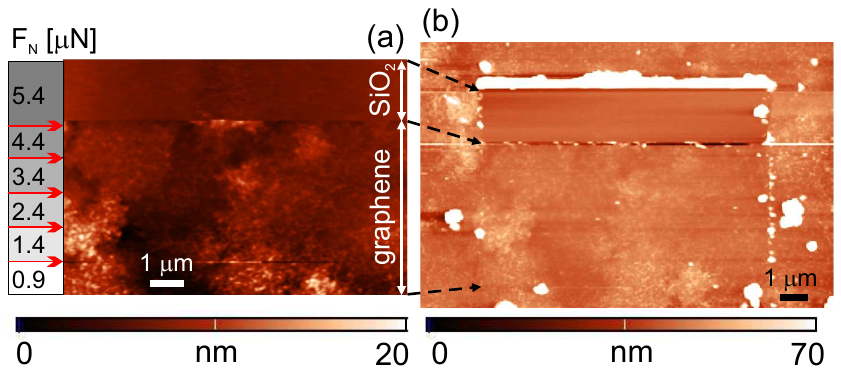}}
	\caption{Wear test: (a) topography during scratching in AFM contact mode, and (b) topography of enlarged area measured in tapping AFM mode after scratching. The value of the normal load is presented on the left side of part (a). The arrows in part (a) mark positions where the normal load was increased.}
	\label{fig_scratch_test}
\end{figure}

Wear tests were done by scratching CVD Gr on $\mathrm{SiO_2}$ in contact AFM mode. The results are presented in Figs. \ref{fig_scratch_test}(a) and \ref{fig_scratch_test}(b), that depict topography obtained during scratching and an enlarged topographic image recorded in tapping mode after scratching, respectively. The scratching was done from the bottom to the top. The normal load was increased in a range starting from $0.9 \ \mathrm{\mu N}$ applied to the bottom Gr stripe with $1 \ \mathrm{\mu m}$ width, to $5.4 \ \mathrm{\mu N}$ applied on the top of the image where tearing occured. The points where the normal load was increased are marked by arrows in Fig. \ref{fig_scratch_test}(a). When the normal load reached a threshold value of $5.4 \ \mathrm{\mu N}$, Gr started to tear. The moment of tearing is clearly visible as a sudden change in the contrast of both topographic images in Fig. \ref{fig_scratch_test}. The normal load was then kept at a high value, while Gr was peeled off by the AFM probe within the scan area. The area where Gr was peeled off is apparent in Fig. \ref{fig_scratch_test}(b) as a rectangular stripe with bare $\mathrm{SiO_2}$. Peeled Gr was rolled and deposited at the top of this domain, where scratching was stopped, and it is visible as a bright and narrow horizontal stripe. 

The same wear experiment was repeated on five different areas of the Gr sample. The results were similar in all cases - sudden Gr tearing at a high enough normal load, while the threshold normal force needed for Gr tearing varied in the range $3.4-5.4 \ \mathrm{\mu N}$. The mechanism of Gr tearing can be explained in the following way. High normal forces applied by the AFM tip during wear test lead to plastic deformations of Gr beneath the tip \cite{vasic_wear}. The plastic deformations are characterized with various defects, mostly by vacancy defects, which degrade the mechanical properties of Gr and its breaking strength \cite{Koratkar_NatCom}. By increasing normal force, Gr becomes more defective while the breaking strength of Gr becomes very small, which results in Gr fracture and tearing. The average threshold force for the tearing of CVD Gr considered here is around $4 \ \mathrm{\mu N}$ and it is much higher, at least by an order of magnitude, than in CVD Gr grown on copper and transferred on $\mathrm{SiO_2}$, where Gr tearing was always initiated from long and wide, thermally grown wrinkles, for normal loads less than $0.5 ~ \mathrm{\mu N}$ \cite{vasic_cvd_gr_carbon} and sometimes already at around $100 ~ \mathrm{nN}$ \cite{Pugno_ACS_2018}. Although in the former cases single-layer Gr samples were considered, the wear resistance of CVD Gr grown on Mo seems to be higher because of the different type of wrinkles in CVD Gr grown on Mo. Here they are small and narrow (simple ripples \cite{Avouris_NL}) and can be easily pressed by the AFM tip without tearing, while the Gr sheet is simultaneously just locally flattened. 

\begin{figure}
	\centerline{\includegraphics[width=8.5cm]{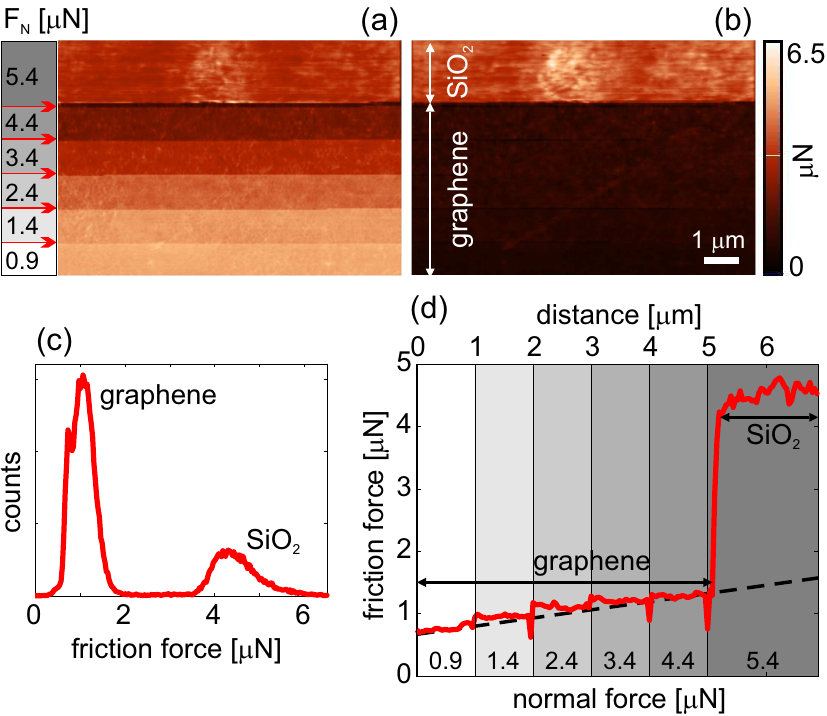}}
	\caption{Friction during the wear test: (a) lateral force map and (b) corresponding friction map during the scratching experiment from Fig. \ref{fig_scratch_test}, (c) histogram of the friction map, and (d) the average vertical profile of the friction force map. The averaging was applied in order to filter out noise and better present stepwise increase of the friction force. The dashed line in part (d) is a liner fit to the stepwise increasing friction force during the scratching of Gr. The slope of this curve corresponds to the friction coefficient of Gr.}
	\label{fig_friction}
\end{figure}

The lateral force recorded during the scratching test is displayed in Fig. \ref{fig_friction}(a). The force increases with the normal load in stepwise fashion before Gr tearing. The friction force was calculated according to the lateral force recorded in forward and backward directions. The friction map is depicted in Fig. \ref{fig_friction}(b), whereas the corresponding histogram is presented in Fig. \ref{fig_friction}(c). The friction map is characterized by two distinct domains: the bottom part with low friction on Gr covered $\mathrm{SiO_2}$ and the top part with high friction on bare $\mathrm{SiO_2}$. According to the histogram, friction on Gr is more than 4 times smaller than on $\mathrm{SiO_2}$, indicating good lubrication properties of Gr. 

The average profile of the friction map along the vertical direction is given in Fig. \ref{fig_friction}(d). The friction force increases stepwise with the normal load. When the threshold force for Gr tearing is reached, the friction increases abruptly and stays at a constant level representing friction on bare $\mathrm{SiO_2}$. The initial stepwise increase of the friction can be approximated by a linear function represented by the dashed line in Fig. \ref{fig_friction}(d). The slope of this linear curve is the ratio between the friction force and applied normal load and it yields a friction coefficient of Gr of only 0.13. This value is similar to that obtained earlier for Gr grown on copper and nickel \cite{Lee_ACSnano}. Friction is also influenced by defects in Gr \cite{Li_MatToday_rev_2019}, mainly by exposed Gr edges and wrinkles, which lead to increased friction \cite{Pugno_ACS_2018, Vasic_gr_edges}. As mentioned above, the considered CVD graphene is almost free of cracks and exposed edges. At the same time, friction maps (the typical one shown in Fig. \ref{fig_friction}(b)) do not show increased friction due to Gr wrinkles, because they are small and narrow and could be easily pressed by the AFM tip. According to these results, CVD Gr grown on Mo could be an excellent choice for making large-scale and ultrathin solid lubricants with increased wear resistance for friction \cite{Lee_ACSnano, graphene_lubricant, Carpick_cvd} and wear reduction \cite{vasic_wear, Won_carbon, Berman_adv_func_mat, friction_reduction_Moseler} of underlying substrates. 

\subsection{Electrical surface potential}

\subsubsection{CVD Gr on Mo}

\begin{figure*}
\centerline{\includegraphics[width=16cm]{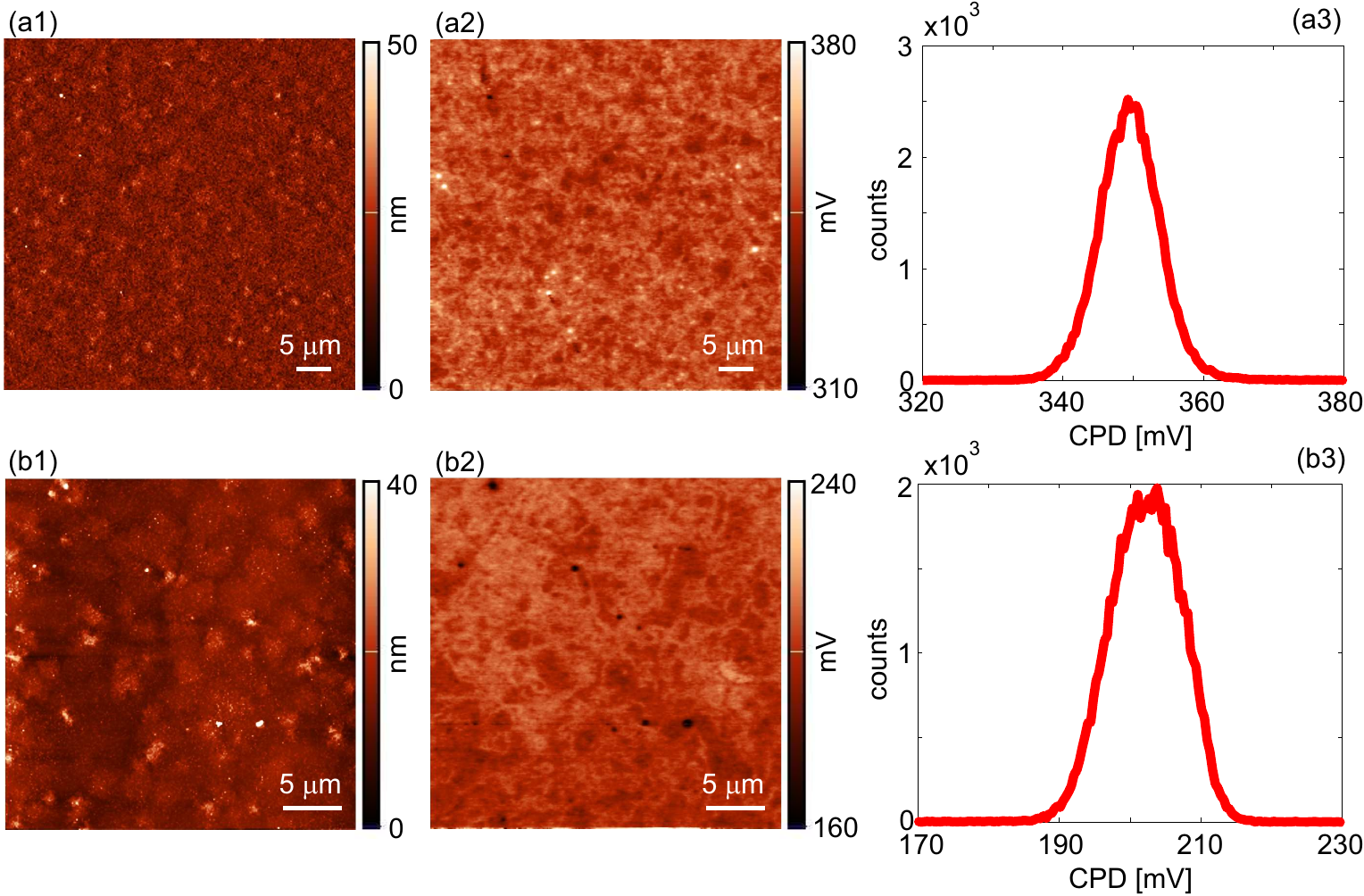}}
\caption{(a1) Morphology of CVD Gr on Mo, (a2) the corresponding CPD map measured by KPFM, and (a3) the histogram of the CPD map. (b1) Morphology of CVD Gr transferred on $\mathrm{SiO_2}$, (b2) the corresponding CPD map measured by KPFM, and (b3) the histogram of the CPD map.}
\label{fig_kpfm}
\end{figure*}

Homogenity of electrical surface potential was investigated by KPFM. Topography of CVD Gr on Mo and the corresponding CPD map are given in Figs. \ref{fig_kpfm}(a1) and \ref{fig_kpfm}(a2), respectively, whereas the histogram of the CPD map is shown in Fig. \ref{fig_kpfm}(a3). The histogram contains a single, narrow peak indicating that the measured CPD is rather uniform over a wide $50 \times 50 \ \mathrm{\mu m^2}$ area. Averaged CPD (taking into account 10 different areas) was $352\pm6 \ \mathrm{mV}$. The absolute value of the work function of the considered CVD Gr on Mo is thus $4.66 \ \mathrm{eV}$. The maximal half-width of all CPD maps measured on $50 \times 50 \ \mathrm{\mu m^2}$ areas was only around $5 \ \mathrm{mV}$, indicating a very uniform electrical surface potential distribution. Still, CPD maps clearly show irregularly shaped potential puddles. The potential between adjacent puddles varies by several $\mathrm{mV}$, while their lateral shapes can not be related to any morphological features such as patches with increased height. Similar electron-hole puddles have been already observed in graphene \cite{martinj_puddles, zhang_puddles, martinsc_puddles, deshpande_puddles} due to charge impurities in the substrate, intercalated between Gr and the substrate, or due to intrinsic ripples in Gr. 

\subsubsection{CVD Gr on $\mathrm{SiO_2}$}

Similar analysis of the distribution of electrical surface potential was done for CVD Gr transferred on $\mathrm{SiO_2}$. The results are given in Fig. \ref{fig_kpfm}(b) representing $30 \times 30 \ \mathrm{\mu m^2}$ topographic and KPFM images (parts (b1) and (b2), respectively), and the CPD histogram (part (b3)). CPD maps exhibit similar features as in the previous case, with a very flat surface potential, implying that CVD Gr on $\mathrm{SiO_2}$ is electrically homogeneous. Since wrinkles are narrow and small, Gr is free of wrinkle-induced potential variations previously observed in other forms of CVD Gr \cite{vasic_cvd_gr_carbon, Ladak, wrinkles_kpfm}. However, small and irregular charge puddles are still present, as in the case on Mo. The average CPD was $205\pm4 \ \mathrm{mV}$, thus giving the absolute value of the work function of Gr transferred on $\mathrm{SiO_2}$ of $4.8 \ \mathrm{eV}$. Therefore, there was a small difference of around $0.14 \ \mathrm{eV}$ between the work functions of Gr on Mo and $\mathrm{SiO_2}$. In the former case, Gr was most probably not thick enough to completely screen an electric field originating from the underlying Mo with a lower work function than Gr. As a result, the work function of Gr on Mo was slightly decreased. A second possibility is that Mo dopes the Gr by charge transfer, again lowering its work function \cite{giovannetti_prl, matkovic_jap}.

The work function of a material or surface is a key property that determines its behavior in an electronic circuit. Energy level differences between different constituent layers of a device dictate functionality ranging from Ohmic contacts to Schottky barriers. One of the primary strengths of silicon and other materials of choice in the semiconductor industry is their uniform work function, or surface potential. Bare silicon surfaces, typically used as references for KPFM measurements, have RMS uniformity on the order of $3 \ \mathrm{mV}$ \cite{leung_nl_2009}. Aside from their use in integrated electronics, surfaces with flat topography and surface potential are also of interest as substrates for self-assembly. The quality of molecular self-assembly is critically determined by the electronic structure of the substrate surface and by variations of its surface potential due to charge transfer between the substrate and adsorbed molecules. As a result, highly homogeneous metal surfaces are often the substrate of choice due to well-defined molecule-metal interactions \cite{Bartels_2010, wyrick_nl_2011, deloach_jchemphys_2017, Schiffrin_2018}. 

Here we show that few-layer Gr grown by CVD on sputtered Mo films has an extremely uniform surface potential profile over large areas, as measured by KPFM. In addition, such Gr that has been transferred, keeps the excellent uniformity, with RMS variability in surface potential in the order of 4 mV for areas as large as $30 \times 30 \ \mathrm{\mu m^2}$. This is an improvement compared to the epitaxial Gr grown on SiC which has surface potential with RMS uniformity on the order of $\sim 10 \ \mathrm{mV}$, however domains of few-layer Gr and steps in the SiC spoil this homogeneity in the surface potential at scales larger than $1 \ \mathrm{\mu m}$ \cite{kazakova_2013}. Monolayer Gr grown by CVD on copper contains wrinkles that also introduce inhomogeneity in surface potential on the order of $\sim 20 \ \mathrm{mV}$ \cite{vasic_cvd_gr_carbon}. Even metal films, such as sputtered gold, display larger variability of surface potential over large areas \cite{hormeno_nanotech_2013}, whereas films deposited with atomic layer deposition display variability on the order of $\sim 10 \ \mathrm{mV}$ \cite{huang_nresl_2017}. 
	
\subsection{Electrical conductivity}

\subsubsection{CVD Gr on Mo}

\begin{figure}
\centerline{\includegraphics[width=8.5cm]{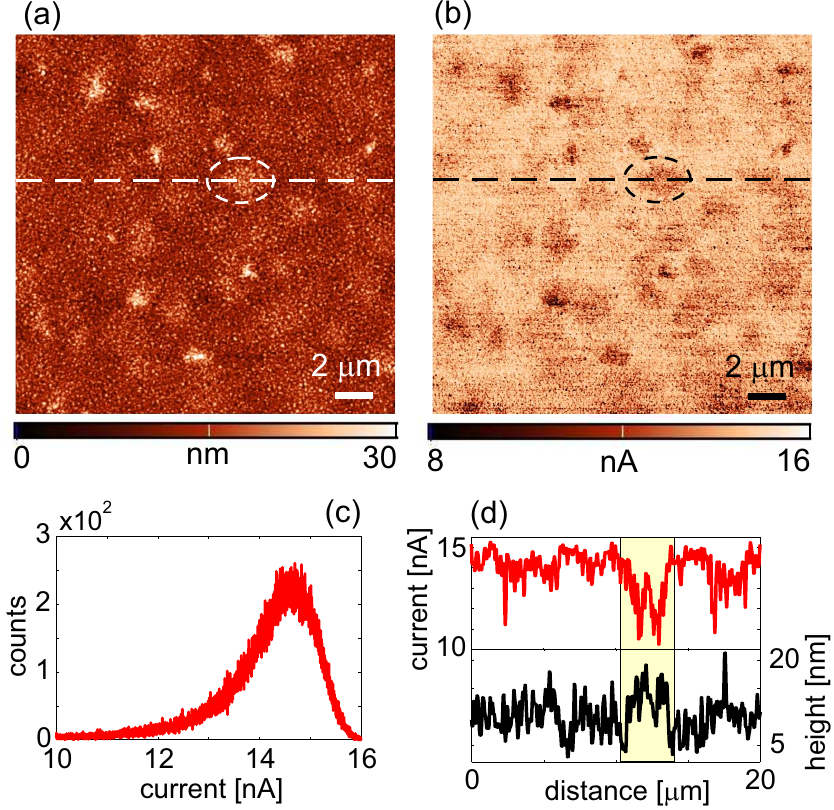}}
\caption{(a) Morphology of CVD Gr on Mo and (b) corresponding current map measured by C-AFM. (c) Histogram of the current distributions in map (b). (d) Height and current profiles along the dashed lines in parts (a) and (b), respectively. Dashed circles in (a) and (b) correspond to the region with lower current which is highlighted by yellow in (d).}
\label{fig_cafm_1}
\end{figure}

Homogenity of electrical conductivity was studied by C-AFM. Topography and corresponding current maps are presented in Figs. \ref{fig_cafm_1}(a) and \ref{fig_cafm_1}(b), respectively. The current map exhibits rather homogeneous and high current. The corresponding histogram is displayed in Fig. \ref{fig_cafm_1}(c). As can be seen, the current distribution is characterized with a single peak around  $14.5 \ \mathrm{nA}$, with a half-width of around $1 \ \mathrm{nA}$. The broadening of the current peak appears due to decreased conductivity on the patches with increased thickness. The height and current profiles across one such patch (along the dashed lines indicated in Figs. \ref{fig_cafm_1}(a) and \ref{fig_cafm_1}(b)) are depicted in Fig. \ref{fig_cafm_1}(d). As can be seen from these profiles, the current drops by several nA on the patch. In addition, small current drops (shown in Fig. S2 of Supplementary material) are visible along narrow Mo grain boundaries (imprinted in Gr as well) because of unstable contact with the AFM tip.

\subsubsection{CVD Gr on $\mathrm{SiO_2}$}

As in the previous case for the electrical surface potential, a similar analysis of the current distribution was conducted for CVD Gr on $\mathrm{SiO_2}$. The conductivity exhibits the same characteristics as previously observed for Gr on Mo: homogeneous and high current except on thicker patches (results presented in Fig. S3 of Supplementary material). 

\begin{figure*}
\centerline{\includegraphics[width=8.5cm]{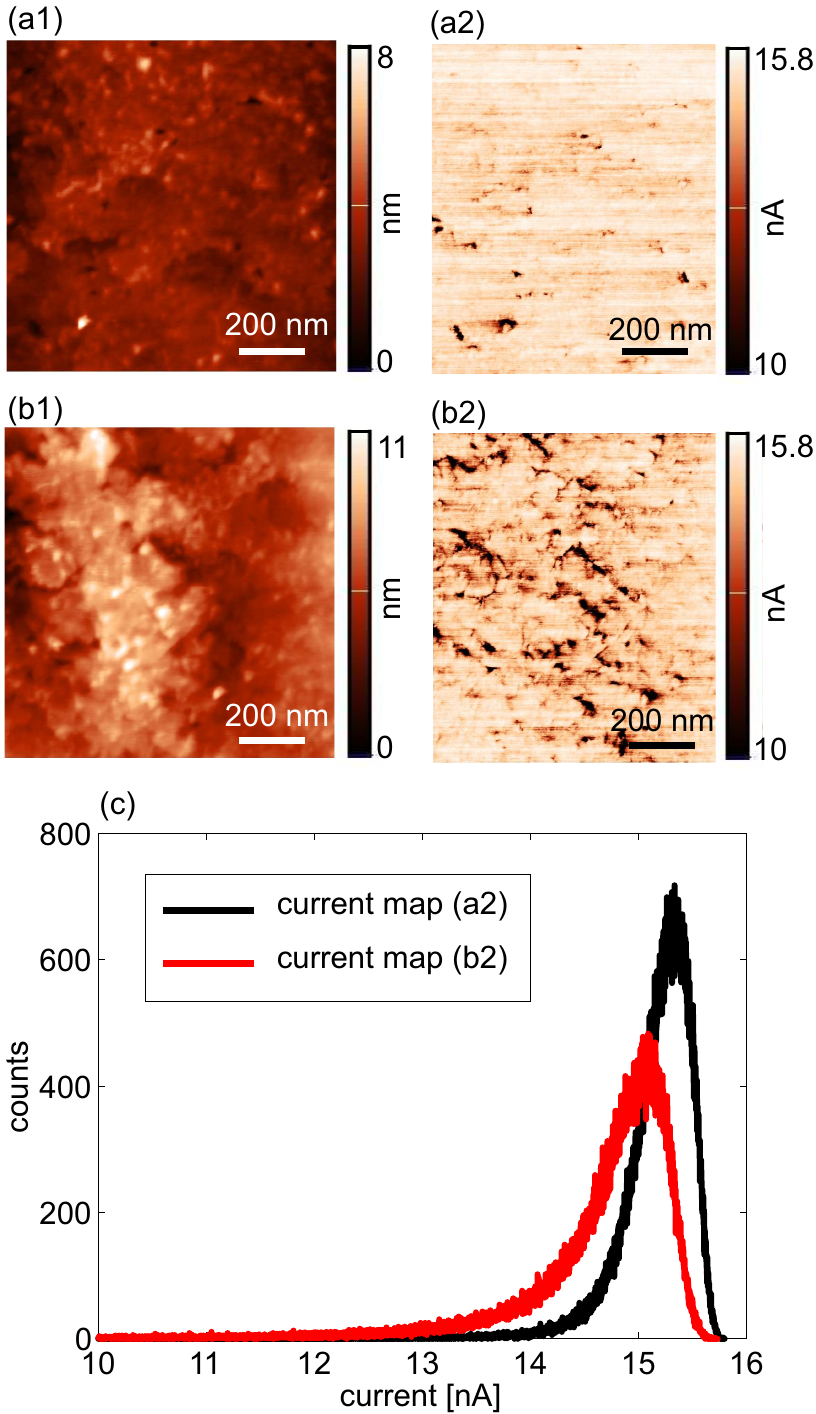}}
\caption{(a1) Morphology and (a2) current map of CVD Gr transferred on $\mathrm{SiO_2}$ without patches. (b1) Morphology and (b2) current map of Gr with a patch. (c) Histograms of the current distributions from parts (a2) and (b2).}
\label{fig_cafm_2}
\end{figure*}

The small-scale images with topographic and current maps are presented for two cases: Figs. \ref{fig_cafm_2}(a1) and \ref{fig_cafm_2}(a2) for flat Gr (without patches) and Figs. \ref{fig_cafm_2}(b1) and \ref{fig_cafm_2}(b2) across a Gr patch. Both current images show homogeneous current despite of a dense network of Gr wrinkles. As we discussed above, wrinkles in the considered case have the geometry of simple, small and narrow ripples \cite{Avouris_NL}. Then, during scanning in AFM contact mode, such wrinkles are pressed by the AFM tip leading to local Gr flattening, which finally gives a constant and high current. At the same time, Gr is wear resistive, so this local mechanical deformation does not result in Gr tearing. This is a significant improvement compared to CVD Gr grown on copper, where wrinkles are much wider and folded \cite{Avouris_NL}, leading to a more pronounced current drop. In the worst case, an AFM tip going across such wrinkles easily initiates local Gr tearing thus producing narrow and insulating trenches in a Gr sheet with zero current \cite{vasic_cvd_gr_carbon}.

Still, the current map in Fig. \ref{fig_cafm_2}(b2) contains local, point-like domains, represented by dark contrast, with a slightly decreased current. Reduced conductivity across patches indicates possible irregularities in the growth of these additional layers. However, the current drops only along patch edges or at pronounced topographic features such as point-like bumps. Therefore, one possible reason of decreased current on these parts is a less stable electrical contact between the AFM tip and Gr. Further improvement in the Gr growth process is needed in order to avoid these imperfections. Current histograms for both Gr with and without patches are given in Fig. \ref{fig_cafm_2}(c). As can be seen, due to a slightly decreased conductivity of the domain with patches, the current peak is shifted by around $0.2 \ \mathrm{nA}$ to a lower value.   

\section{Conclusion}

In summary, we have demonstrated that although few-layer Gr grown on Mo does contain wrinkles with a height of several nanometers, the wrinkles are much narrower than in CVD Gr grown on copper, and they do not have a detrimental effect on uniformity of wear and electrical properties. It is shown that few-layer Gr grown by CVD on sputtered Mo films is characterized with a very low friction coefficient of around 0.13. Its wear resistance is improved compared to CVD Gr grown on copper, giving the threshold normal load for wear of around $4 \ \mathrm{\mu N}$. The considered Gr has very uniform surface potential over large areas, with RMS variability on the order of $5 \ \mathrm{mV}$ for areas as large as $50 \times 50 \ \mu m^2$. The uniformity of electrical properties is better than in other types of Gr and is on par with industrial-grade materials such as silicon and metals deposited by atomic layer deposition. The local conductivity of the Gr films is also uniform, although with small variations at the edges of Gr patches with varying thickness. The patches are a result of the growth process which should be further optimized in order to overcome this issue. 

We thus propose that few-layer Gr grown on Mo holds strong potential for use as an ultrathin solid lubricant for friction and wear reduction. It can also be used as an ultrathin electrode in integrated electronics, allowing wafer-scale device uniformity and reproducibility. Furthermore, the material holds potential as a substrate for self-assembly and for other uses that require uniform and well-defined electrical properties over large areas.

We would like to acknowledge support of the Ministry of Education, Science, and Technological Development of the Republic of Serbia through projects OI171005 and TR32008. We acknowledge the Innovation Fund project no 50038 and the Graphene Flagship.

\section*{References}


\begin{thebibliography}{10}
	\expandafter\ifx\csname url\endcsname\relax
	\def\url#1{\texttt{#1}}\fi
	\expandafter\ifx\csname urlprefix\endcsname\relax\def\urlprefix{URL }\fi
	\expandafter\ifx\csname href\endcsname\relax
	\def\href#1#2{#2} \def\path#1{#1}\fi
	
	\bibitem{Li_Science}
	X.~Li, W.~Cai, J.~An, S.~Kim, J.~Nah, D.~Yang, R.~Piner, A.~Velamakanni,
	I.~Jung, E.~Tutuc, S.~K. Banerjee, L.~Colombo, R.~S. Ruoff, {Large-Area
		Synthesis of High-Quality and Uniform Graphene Films on Copper Foils},
	Science 324 (2009) 1312--1314 (2009).
	
	\bibitem{Obraztsov}
	A.~N. Obraztsov, E.~A. Obraztsova, A.~V. Tyurnina, A.~A. Zolotukhin, Chemical
	vapor deposition of thin graphite films of nanometer thickness, Carbon 45
	(2007) 2017--2021 (2007).
	
	\bibitem{zhou_nat_comm}
	H.~Zhou, W.~J. Yu, L.~Liu, R.~Cheng, Y.~Chen, X.~Huang, Y.~Liu, Y.~Wang,
	Y.~Huang, X.~Duan, Chemical vapour deposition growth of large single crystals
	of monolayer and bilayer graphene, Nat. Comm. 4 (2013) 2096 (2013).
	
	\bibitem{Kim_Nature}
	K.~S. Kim, Y.~Zhao, H.~Jang, S.~Y. Lee, J.~M. Kim, K.~S. Kim, J.-H. Ahn,
	P.~Kim, J.-Y. Choi, B.~H. Hong, Large-scale pattern growth of graphene films
	for stretchable transparent electrodes, Nature 457 (2009) 706--710 (2009).
	
	\bibitem{Li_NL}
	X.~Li, Y.~Zhu, W.~Cai, M.~Borysiak, B.~Han, D.~Chen, R.~D. Piner, L.~Colombo,
	R.~S. Ruoff, {Transfer of Large-Area Graphene Films for High-Performance
		Transparent Conductive Electrodes}, Nano Lett. 9 (2009) 4359--4363 (2009).
	
	\bibitem{transfer_nanoscale}
	J.~Kang, D.~Shin, S.~Bae, B.~H. Hong, Graphene transfer: key for applications,
	Nanoscale 4 (2012) 5527--5537 (2012).
	
	\bibitem{review_jmchem}
	C.~Mattevi, H.~Kim, M.~Chhowalla, A review of chemical vapour deposition of
	graphene on copper, J. Mater. Chem. 21 (2011) 3324--3334 (2011).
	
	\bibitem{review_chemres}
	Y.~Zhang, L.~Zhang, C.~Zhou, {Review of Chemical Vapor Deposition of Graphene
		and Related Applications}, Acc. Chem. Res. 46 (2013) 2329--2339 (2013).
	
	\bibitem{review_carbon}
	C.-M. Seah, S.-P. Chai, A.~R. Mohamed, Mechanisms of graphene growth by
	chemical vapour deposition on transition metals, Carbon 70 (2014) 1 -- 21
	(2014).
	
	\bibitem{RuizVargas_NL}
	C.~S. Ruiz-Vargas, H.~L. Zhuang, P.~Y. Huang, A.~M. van~der Zande, S.~Garg,
	P.~L. McEuen, D.~A. Muller, R.~G. Hennig, J.~Park, {Softened Elastic Response
		and Unzipping in Chemical Vapor Deposition Graphene Membranes}, Nano Lett. 11
	(2011) 2259--2263 (2011).
	
	\bibitem{Lee_Science}
	G.-H. Lee, R.~C. Cooper, S.~J. An, S.~Lee, A.~van~der Zande, N.~Petrone, A.~G.
	Hammerberg, C.~Lee, B.~Crawford, W.~Oliver, J.~W. Kysar, J.~Hone,
	{High--Strength Chemical--Vapor–-Deposited Graphene and Grain Boundaries},
	Science 340 (2013) 1073--1076 (2013).
	
	\bibitem{Rasool_NatCom}
	H.~I. Rasool, C.~Ophus, W.~S. Klug, A.~Zettl, J.~K. Gimzewski, Measurement of
	the intrinsic strength of crystalline and polycrystalline graphene, Nat.
	Commun. 4 (2013) 2811 (2013).
	
	\bibitem{Koepke_NL}
	J.~C. Koepke, J.~D. Wood, D.~Estrada, Z.-Y. Ong, K.~T. He, E.~Pop, J.~W.
	Lyding, {Atomic-Scale Evidence for Potential Barriers and Strong Carrier
		Scattering at Graphene Grain Boundaries: A Scanning Tunneling Microscopy
		Study}, Nano Lett. 7 (2013) 75--86 (2013).
	
	\bibitem{Clark_NL}
	K.~W. Clark, X.-G. Zhang, I.~V. Vlassiouk, G.~He, R.~M. Feenstra, A.-P. Li,
	{Spatially Resolved Mapping of Electrical Conductivity across Individual
		Domain (Grain) Boundaries in Graphene}, Nano Lett. 7 (2013) 7956--7966
	(2013).
	
	\bibitem{NemesIncze_carbon}
	P.~Nemes-Incze, P.~Vancs\'o, Z.~Osv\'ath, G.~I. M\'ark, X.~Jin, Y.-S. Kim,
	C.~Hwang, P.~Lambin, C.~Chapelier, L.~B. Bir\'o, Electronic states of
	disordered grain boundaries in graphene prepared by chemical vapor
	deposition, Carbon 64 (2013) 178--186 (2013).
	
	\bibitem{Roche_AdvMat}
	A.~W. Cummings, D.~L. Duong, V.~L. Nguyen, D.~Van~Tuan, J.~Kotakoski, J.~E.
	Barrios~Vargas, Y.~H. Lee, S.~Roche, {Charge Transport in Polycrystalline
		Graphene: Challenges and Opportunities}, Adv. Mater. 26 (2014) 5079--5094
	(2014).
	
	\bibitem{Xu_NL}
	K.~Xu, P.~Cao, J.~R. Heath, {Scanning Tunneling Microscopy Characterization of
		the Electrical Properties of Wrinkles in Exfoliated Graphene Monolayers},
	Nano Lett. 9 (2009) 4446--4451 (2009).
	
	\bibitem{Liu_NanoResearch}
	N.~Liu, Z.~Pan, L.~Fu, C.~Zhang, B.~Dai, Z.~Liu, The origin of wrinkles on
	transferred graphene, Nano Res. 4 (2011) 996--1004 (2011).
	
	\bibitem{Ahmad_Nanotech}
	M.~Ahmad, H.~An, Y.~S. Kim, J.~H. Lee, J.~Jung, S.-H. Chun, Y.~Seo, Nanoscale
	investigation of charge transport at the grain boundaries and wrinkles in
	graphene film, Nanotechnology 23 (2012) 285705 (2012).
	
	\bibitem{Avouris_NL}
	W.~Zhu, T.~Low, V.~Perebeinos, A.~A. Bol, Y.~Zhu, H.~Yan, J.~Tersoff,
	P.~Avouris, {Structure and Electronic Transport in Graphene Wrinkles}, Nano
	Lett. 12 (2012) 3431--3436 (2012).
	
	\bibitem{Ladak}
	S.~Ladak, J.~M. Ball, D.~Moseley, G.~Eda, W.~R. Branford, M.~Chhowalla, T.~D.
	Anthopoulos, L.~F. Cohen, Observation of wrinkle induced potential drops in
	biased chemically derived graphene thin film networks, Carbon 64 (2013)
	35--44 (2013).
	
	\bibitem{wrinkles_kpfm}
	P.~Willke, C.~M\"ohle, A.~Sinterhauf, T.~Kotzott, H.~K. Yu, A.~Wodtke,
	M.~Wenderoth, {Local transport measurements in graphene on $\mathrm{SiO_2}$
		using Kelvin probe force microscopy}, Carbon 102 (2016) 470 -- 476 (2016).
	
	\bibitem{vasic_cvd_gr_carbon}
	B.~Vasi\'c, A.~Zurutuza, R.~Gaji\'c, Spatial variation of wear and electrical
	properties across wrinkles in chemical vapour deposition graphene, Carbon 102
	(2016) 304 -- 310 (2016).
	
	\bibitem{synthesis_carbon}
	Y.~Wu, G.~Yu, H.~Wang, B.~Wang, Z.~Chen, Y.~Zhang, B.~Wang, X.~Shi, X.~Xie,
	Z.~Jin, X.~Liu, Synthesis of large-area graphene on molybdenum foils by
	chemical vapor deposition, Carbon 50 (2012) 5226 -- 5231 (2012).
	
	\bibitem{synthesis_sten}
	Y.~Grachova, S.~Vollebregt, A.~L. Lacaita, P.~M. Sarro, {High Quality
		Wafer-scale CVD Graphene on Molybdenum Thin Film for Sensing Application},
	Procedia Eng. 87 (2014) 1501 -- 1504 (2014).
	
	\bibitem{synthesis_NL_paper}
	Z.~Zou, L.~Fu, X.~Song, Y.~Zhang, Z.~Liu, {Carbide-Forming Groups IVB-VIB
		Metals: A New Territory in the Periodic Table for CVD Growth of Graphene},
	Nano Lett. 14 (2014) 3832--3839 (2014).
	
	\bibitem{gas_sensor2}
	C.~Schiattarella, S.~Vollebregt, T.~Polichetti, B.~Alfano, E.~Massera, M.~L.
	Miglietta, G.~Di~Francia, P.~M. Sarro, {CVD transfer-free graphene for
		sensing applications}, Beilstein J. Nanotechnol. 8 (2017) 1015–1022 (2017).
	
	\bibitem{pressure_sensing}
	S.~Vollebregt, R.~J. Dolleman, H.~S.~J. van~der Zant, P.~G. Steeneken, P.~M.
	Sarro, Suspended graphene beams with tunable gap for squeeze-film pressure
	sensing, in: 2017 19th International Conference on Solid-State Sensors,
	Actuators and Microsystems (TRANSDUCERS), 2017, pp. 770--773 (2017).
	
	\bibitem{corrosion1}
	S.~Naghdi, I.~Jevremovi\'c, V.~Mi\v{s}kovi\'c-Stankovi\'c, K.~Y. Rhee,
	{Chemical vapour deposition at atmospheric pressure of graphene on molybdenum
		foil: Effect of annealing time on characteristics and corrosion stability of
		graphene coatings}, Corros. Sci. 113 (2016) 116 -- 125 (2016).
	
	\bibitem{corrosion2}
	S.~Naghdi, K.~Ne\v{s}ovi\'c, V.~Mi\v{s}kovi\'c-Stankovi\'c, K.~Y. Rhee,
	Comprehensive electrochemical study on corrosion performance of graphene
	coatings deposited by chemical vapour deposition at atmospheric pressure on
	platinum-coated molybdenum foil, Corros. Sci. 130 (2018) 31 -- 44 (2018).
	
	\bibitem{oxidation}
	S.~Naghdi, K.~Y. Rhee, S.~J. Park, Oxidation resistance of graphene-coated
	molybdenum: Effects of pre-washing and hydrogen flow rate, Int. J. Refract.
	Metals Hard Mater. 65 (2017) 29 -- 33 (2017).
	
	\bibitem{gas_sensor1}
	F.~Ricciardella, S.~Vollebregt, T.~Polichetti, M.~Miscuglio, B.~Alfano, M.~L.
	Miglietta, E.~Massera, G.~Di~Francia, P.~M. Sarro, Effects of graphene
	defects on gas sensing properties towards $\mathrm{NO_2}$ detection,
	Nanoscale 9 (2017) 6085--6093 (2017).
	
	\bibitem{lf_calibraction}
	D.~K. Hong, S.~A. Han, J.~H. Park, S.~H. Tan, N.~Lee, Y.~Seo, {Frictional force
		detection from lateral force microscopic image using a Si grating}, Colloids
	Surf. A 313-314 (2008) 567 -- 570 (2008).
	
	\bibitem{kpfm_calibration}
	Y.-J. Yu, Y.~Zhao, S.~Ryu, L.~E. Brus, K.~S. Kim, P.~Kim, {Tuning the graphene
		work function by electric field effect}, Nano Lett. 9 (2009) 3430--4 (2009).
	
	\bibitem{graf_nl_20107}
	D.~Graf, F.~Molitor, K.~Ensslin, C.~Stampfer, A.~Jungen, C.~Hierold, L.~Wirtz,
	{Spatially Resolved Raman Spectroscopy of Single- and Few-Layer Graphene},
	Nano Lett. 7 (2007) 238--242 (2007).
	
	\bibitem{vasic_wear}
	B.~Vasi\'c, A.~Matkovi\'c, U.~Ralevi\'c, M.~Beli\'c, R.~Gaji\'c, Nanoscale wear
	of graphene and wear protection by graphene, Carbon 120 (2017) 137 -- 144
	(2017).
	
	\bibitem{Koratkar_NatCom}
	A.~Zandiatashbar, G.-H. Lee, S.~J. An, S.~Lee, N.~Mathew, M.~Terrones,
	T.~Hayashi, C.~R. Picu, J.~Hone, N.~Koratkar, Effect of defects on the
	intrinsic strength and stiffness of graphene, Nat. Commun. 5 (2014) 3186
	(2014).
	
	\bibitem{Pugno_ACS_2018}
	M.~Tripathi, F.~Awaja, R.~A. Bizao, S.~Signetti, E.~Iacob, G.~Paolicelli,
	S.~Valeri, A.~Dalton, N.~M. Pugno, {Friction and Adhesion of Different
		Structural Defects of Graphene}, ACS Appl. Mater. Inter. 10 (2018)
	44614--44623 (2018).
	
	\bibitem{Lee_ACSnano}
	K.-S. Kim, H.-J. Lee, C.~Lee, S.-K. Lee, H.~Jang, J.-H. Ahn, J.-H. Kim, H.-J.
	Lee, {Chemical Vapor Deposition-Grown Graphene: The Thinnest Solid
		Lubricant}, ACS Nano 5 (2011) 5107--5114 (2011).
	
	\bibitem{Li_MatToday_rev_2019}
	S.~Zhang, T.~Ma, A.~Erdemir, Q.~Li, {Tribology of two-dimensional materials:
		From mechanisms to modulating strategies}, Mater. Today 26 (2019) 67 -- 86
	(2019).
	
	\bibitem{Vasic_gr_edges}
	A.~Vasi\'c, B.~Matkovi\'c, R.~Gaji\'c, I.~Stankovi\'c, Wear properties of
	graphene edges probed by atomic force microscopy based lateral manipulation,
	Carbon 107 (2016) 723--732 (2016).
	
	\bibitem{graphene_lubricant}
	D.~Berman, A.~Erdemir, A.~V. Sumant, Graphene: a new emerging lubricant, Mater.
	Today 17 (2014) 31--42 (2014).
	
	\bibitem{Carpick_cvd}
	P.~Egberts, G.~H. Han, X.~Z. Liu, A.~T.~C. Johnson, R.~W. Carpick, {Frictional
		Behavior of Atomically Thin Sheets: Hexagonal-Shaped Graphene Islands Grown
		on Copper by Chemical Vapor Deposition}, ACS Nano 8 (2014) 5010--5021 (2014).
	
	\bibitem{Won_carbon}
	M.-S. Won, O.~V. Penkov, D.-E. Kim, {Durability and degradation mechanism of
		graphene coatings deposited on Cu substrates under dry contact sliding},
	Carbon 54 (2013) 472--481 (2013).
	
	\bibitem{Berman_adv_func_mat}
	D.~Berman, S.~A. Deshmukh, S.~K. R.~S. Sankaranarayanan, A.~Erdemir, A.~V.
	Sumant, {Extraordinary Macroscale Wear Resistance of One Atom Thick Graphene
		Layer}, Adv. Funct. Mater. 24 (2014) 6640--6646 (2014).
	
	\bibitem{friction_reduction_Moseler}
	A.~Klemenz, L.~Pastewka, S.~G. Balakrishna, A.~Caron, R.~Bennewitz, M.~Moseler,
	{Atomic Scale Mechanisms of Friction Reduction and Wear Protection by
		Graphene}, Nano Lett. 14 (2014) 7145--7152 (2014).
	
	\bibitem{martinj_puddles}
	J.~Martin, N.~Akerman, G.~Ulbricht, T.~Lohmann, J.~H. Smet, K.~von Klitzing,
	A.~Yacoby, {Observation of electron-hole puddles in graphene using a scanning
		single-electron transistor}, Nat. Phys. 4 (2007) 144 (2007).
	
	\bibitem{zhang_puddles}
	Y.~Zhang, V.~W. Brar, C.~Girit, A.~Zettl, M.~F. Crommie, {Origin of spatial
		charge inhomogeneity in graphene}, Nat. Phys. 5 (2009) 722 (2009).
	
	\bibitem{martinsc_puddles}
	S.~C. Martin, S.~Samaddar, B.~Sac\'ep\'e, A.~Kimouche, J.~Coraux, F.~Fuchs,
	B.~Gr\'evin, H.~Courtois, C.~B. Winkelmann, Disorder and screening in
	decoupled graphene on a metallic substrate, Phys. Rev. B 91 (2015) 041406
	(2015).
	
	\bibitem{deshpande_puddles}
	A.~Deshpande, W.~Bao, F.~Miao, C.~N. Lau, B.~J. LeRoy, {Spatially resolved
		spectroscopy of monolayer graphene on $\mathrm{SiO_2}$}, Phys. Rev. B 79
	(2009) 205411 (2009).
	
	\bibitem{giovannetti_prl}
	G.~Giovannetti, P.~A. Khomyakov, G.~Brocks, V.~M. Karpan, J.~van~den Brink,
	P.~J. Kelly, {Doping Graphene with Metal Contacts}, Phys. Rev. Lett. 101
	(2008) 026803 (2008).
	
	\bibitem{matkovic_jap}
	A.~Matkovi\'c, M.~Chhikara, M.~Mili\'cevi\'c, U.~Ralevi\'c, B.~Vasi\'c,
	D.~Jovanovi\'c, M.~Beli\'c, G.~Bratina, R.~Gaji\'c, Influence of a gold
	substrate on the optical properties of graphene, J. Appl. Phys. 117 (2015)
	015305 (2015).
	
	\bibitem{leung_nl_2009}
	C.~Leung, H.~Kinns, B.~W. Hoogenboom, S.~Howorka, P.~Mesquida, {Imaging Surface
		Charges of Individual Biomolecules}, Nano Lett. 9 (2009) 2769--2773 (2009).
	
	\bibitem{Bartels_2010}
	L.~Bartels, {Tailoring molecular layers at metal surfaces}, Nat. Chem. 2 (2010)
	87--95 (2010).
	
	\bibitem{wyrick_nl_2011}
	J.~Wyrick, D.-H. Kim, D.~Sun, Z.~Cheng, W.~Lu, Y.~Zhu, K.~Berland, Y.~S. Kim,
	E.~Rotenberg, M.~Luo, P.~Hyldgaard, T.~L. Einstein, L.~Bartels, {Do
		Two-Dimensional "Noble Gas Atoms" Produce Molecular Honeycombs at a Metal
		Surface?}, Nano Lett. 11 (2011) 2944–2948 (2011).
	
	\bibitem{deloach_jchemphys_2017}
	A.~S. DeLoach, B.~R. Conrad, T.~L. Einstein, D.~B. Dougherty, {Coverage
		dependent molecular assembly of anthraquinone on Au(111)}, J. Chem. Phys. 147
	(2017) 184701 (2017).
	
	\bibitem{Schiffrin_2018}
	C.~Krull, M.~Castelli, P.~Hapala, D.~Kumar, A.~Tadich, M.~Capsoni, M.~T.
	Edmonds, J.~Hellerstedt, S.~A. Burke, P.~Jelinek, A.~Schiffrin, {Iron-based
		trinuclear metal-organic nanostructures on a surface with local charge
		accumulation}, Nat. Commun. 9 (2018) 3211 (2018).
	
	\bibitem{kazakova_2013}
	O.~Kazakova, V.~Panchal, T.~L. Burnett, {Epitaxial Graphene and Graphene-Based
		Devices Studied by Electrical Scanning Probe Microscopy}, Crystals 3 (2013)
	191--233 (2013).
	
	\bibitem{hormeno_nanotech_2013}
	S.~Horme\~{n}o, M.~Penedo, C.~V. Manzano, M.~Luna, {Gold nanoparticle coated
		silicon tips for Kelvin probe force microscopy in air}, Nanotechnology 24
	(2013) 395701 (2013).
	
	\bibitem{huang_nresl_2017}
	R.~Huang, S.~Ye, K.~Sun, K.~S. Kiang, C.~H.~K. de~Groot, {Fermi Level Tuning of
		ZnO Films Through Supercycled Atomic Layer Deposition}, Nanoscale Res. Lett.
	12 (2017) 541 (2017).
	
\end{thebibliography}

\end{document}